\documentclass[preprint,pre,a4paper,superscriptaddress,longbibliography]{revtex4-1}
\usepackage{graphicx,color,epsfig,subfig,dcolumn,bm,mathrsfs,amsmath,booktabs}
\usepackage[colorlinks]{hyperref}
\usepackage{color}
\makeatletter
\renewcommand{\p@subsection}{}
\renewcommand{\p@subsubsection}{}
\makeatother

\newcommand{\ud}{\mathrm{d}}

\newcommand{\change}[1]{\textcolor{black}{#1}}

\begin{document}

\title{Capillary rising in a tube with corners}

\author{Chen Zhao}
\affiliation{Center of Soft Matter Physics and Its Applications, Beihang University, Beijing 100191, China}
\affiliation{School of Physics, Beihang University, Beijing 100191, China}

\author{Jiajia Zhou}
\email[]{zhouj2@scut.edu.cn}
\affiliation{South China Advanced Institute for Soft Matter Science and Technology, School of Emergent Soft Matter, South China University of Technology, Guangzhou 510640, China}
\affiliation{Guangdong Provincial Key Laboratory of Functional and Intelligent Hybrid Materials and Devices, South China University of Technology, Guangzhou 510640, China}

\author{Masao Doi}
\affiliation{Center of Soft Matter Physics and Its Applications, Beihang University, Beijing 100191, China}
\affiliation{Wenzhou Institute, University of Chinese Academy of Science, 
Wenzhou, Zhejiang 325000, China}

%\date{\today}

\begin{abstract}
We study the dynamics of a fluid rising in a capillary tube with corners.
In the cornered tube, unlike the circular tube, fluid rises with two parts, the bulk part where the entire cross-section is occupied by the fluid, and the finger part where the cross-section is only partially filled.
Using Onsager principle, we derive coupled time-evolution equations for the two parts.  
We show that (a) at the early stage of rising, the dynamics is dominated  by
the bulk part and the fluid height $h_0(t)$ shows the same behavior as that
in the circular tube, and (b) at the late stage, the bulk part stops rising,
but the finger part keeps rising following the scaling law of $h_1(t) \sim t^{1/3}$.
We also show that due to the coupling between the two parts, the equilibrium bulk height is smaller than the Jurin's height which ignores the effect of the finger part.
\end{abstract}

%\pacs{}

\maketitle

%%%%%%%%%%%%%%%%%%%%%%%%%%%%%%%%%%%%%%%%%%%%%%%%%%%%%%%%%%%%%%
%%%%%%%%%%%%%%%%%%%%%%%%%%%%%%%%%%%%%%%%%%%%%%%%%%%%%%%%%%%%%%
\section{Introduction}

When a capillary tube is inserted into a fluid reservoir, the fluid rises inside the capillary and eventually takes an equilibrium height.
Such capillary filling phenomena have been extensively studied since they are related to various applications, such as lithography \cite{Unger2000Monolithic}, DNA translocation \cite{2001Voltage,
Yitzhak2005DNA}, templates for nanoscale self-assembly \cite{2006Solvent}, and microfluidic devices \cite{Gau1999Liquid, 2010A}.

Lucas \cite{Lucas1918} and Washburn \cite{Washburn1921} showed that at an early stage of rising, the filling length $h(t)$ increases with time $t$ following the scaling relation $h(t) \sim t^{1/2}$.
The $t^{1/2}$ scaling law characterizes the imbibition dynamics driven by the surface tension, and has been confirmed in many nanoscopic systems \cite{2007Capillary1,2008Dynamics1, YaoYang2017, YaoYang2018, 2018_square}.
\change{(Other scaling laws have been proposed for tubes that have varied diameters \cite{Reyssat2008, Gorce2016}).}
On the other hand, in macroscopic systems, the $t^{1/2}$ scaling law ceases to be valid when the gravity becomes non-negligible.
In a circular tube of radius $r$, the meniscus height $h(t)$ approaches to an equilibrium height, called Jurin's height \cite{Jurin1718, dBQ}, which is given by $H_{\rm J} = 2 \gamma \cos \theta / \rho g r$, where $\gamma$ is the surface tension of the fluid, $\theta$ is the equilibrium contact angle of the fluid at the capillary, $\rho$ is the density of the fluid, and $g$ is the gravitational constant.

%For the bulk liquid rising, basically thre periods appear in the process. The first period is
% referred as the inertial regime, in which the capillary and inertial forces dominate the
% imbibition, leading to filling length evolves a linear relationship with time:
%$h(t) \sim t$ \cite{2012Early}, and the time scale for the inertial regime is
%$t \sim \sqrt{ \rho R^3 / \gamma }$. As the imbibition length grows a bit longer, the
%inertial force gets negligibly small, and the filling process reaches the viscous
%regime, which the capillary and the viscous forces play an important part in and
%the gravitational force is neglected, resulting the filling length changes with $t^{1/2}$.
%The last evolutionary period is the asymptotic equilibrium regime, which the capillary
%force, the viscous force, and the gravitational force take part in, giving the filling
%length a final state by Jurin's height \cite{dBQ} as
%$h_{Jurin} = 2 \gamma \cos \theta / \rho g a$.

If the cross-section of the tube has corners, the liquid rising does not stop since in the very small region near the corner, \change{the reduction in surface energy is bigger than the increase in gravitational energy}.
Indeed, the equilibrium profile of the meniscus near the corner is hyperbolic \cite{Taylor1710, Hauksbee1710}, and therefore the tip of the finger must keep rising.
Dong and Chatzis \cite{Dong1995} calculated the growth of the finger at the corner of a square tube in the absence of gravity (the situation that the tube is placed horizontally), and showed that the advancement of the finger tip obeys the same scaling law as Lucas-Washburn.

On the other hand, many studies have been conducted for the liquid rise in a corner made by two vertical plates, so-called Taylor rising setup \cite{Taylor1710}.
Such studies have indicated both experimentally and theoretically that in the presence of gravity, the advancement of the finger obeys scaling law different from Lucas-Washburn; the finger length increases in proportion to $t^{1/3}$ \cite{Tang1994, Higuera2008, Ponomarenko2011, Heshmati2014, 2019_Taylor_rising, 2020_onethird}.

In the previous studies on the liquid imbibition in the corner, attention has been focused on the finger part only, and the bulk part has been assumed to be fixed.
If the liquid imbibition takes place in a cornered tube, both bulk part and the finger part
are evolving in time.
In our previous works \cite{2018_square, 2021_rect_LW}, we considered the interaction between the bulk part and the finger part.
We constructed a set of equations which determine the time evolution of both parts, and showed that the interaction can be important.
In those works, however, we considered the case of a horizontal tube (no gravity effect), and therefore the conclusion was somewhat unsurprising, that both the bulk part and the finger part increase obeying the $t^{1/2}$ scaling law, and that the interaction only changes the coefficients.

The dynamics of capillary rising in a square tube has been studied both experimentally \cite{Ouali2013, Heshmati2014, Wijnhorst2020} and numerically \cite{Gurumurthy2018a, ZhaoJianlin2021}.
Gurumurthy \emph{et al.} \cite{Gurumurthy2018a} used volume-of-fluid method with adaptive mesh refinement.
Zhao \emph{et al.} \cite{ZhaoJianlin2021} extended Interacting Capillary Bundle method \cite{DongMingzhe1998, DongMingzhe2005, DongMingzhe2006} to study the corner flow in a square tube.
In this paper, we develop a general theory for the capillary rise in a tube with corners, and use a square tube as an example to demonstrate the applicability of the method.
%The theoretical method is the same as in referece \cite{2018_square}.
Using Onsager variational principle \cite{DoiSoft}, we derive the coupled equations for the bulk part and the finger part, and study the effect of the coupling.
We shall show that the advance of the finger front follows a time-scaling of $t^{1/3}$, and this scaling is universal for general corner flows.
We also show that due to the presence of the finger part, the equilibrium height of the bulk part is always smaller than Jurin's height and that this effect is characterized by a dimensionless parameter $s^*$, called the equilibrium saturation \cite{2021_rect_sstar, 2021_rect_LW}: Bigger $s^*$ is, smaller height the bulk can reach.

%%%%%%%%%%%%%%%%%%%%%%%%%%%%%%%%%%%%%%%%%%%%%%%%%%
\section{Methods}

%\subsection{Time evolution equations}

%-----------------------------------------------------
\begin{figure}[htbp]
  \centering
  \includegraphics[width=0.7\columnwidth]{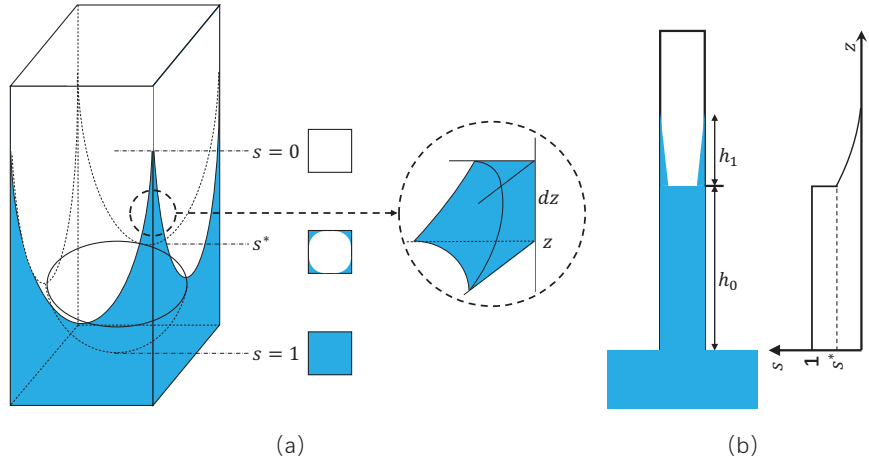}
  \caption{(a) Sketch of capillary rising in a square tube. The liquid inside the tube can be separated into three regions: bulk part with saturation $s=1$; finger part starting with $s=s^*$; and the transition region in between. (b) Our model neglects the transition region. The saturation $s$ changes from 1 to $s^*$ discontinuously at height $h_0$. The length of the finger is denoted by $h_1$. The saturation profile $s(z)$ is also shown.}
  \label{fig:sketch}
\end{figure}
%-----------------------------------------------------

We consider a capillary tube with corners, vertically inserted to a fluid reservoir.
The inner surface of the tube is wetted by the fluid, thus the capillary force drives the liquid up along the tube. 
A schematic picture of the system is illustrated in Fig.~\ref{fig:sketch}(a), using a square tube of side length $2a$ for demonstration. 
\change{Our derivation is based on the assumption that the inertial can be ignored. 
The inertial is important at the initial stage of the capillary rising \cite{Quere1997}.}
The following derivation is general, and it can be applied to any tubes with uniform shape of the cross-section. 

% We consider a capillary of square tube  of side length $2a$ ,
% vertically inserted to a fluid reservoir (see Fig.~\ref{fig1}, ).
% The inner surface of the tube is wetted by the fluid, thus the capillary force drives the
% liquid up along the tube.  Though we consider a square tube, the following discussion
% is general, and can be applied to any tube which has uniform cross-section.

We take the $z$ coordinate in vertical direction (the origin being set at the height of the bulk liquid).
Let $s(z;t)$ be the area fraction occupied by the fluid in the tube at the height $z$ at time $t$: $s(z;t)$ is equal  to 1 at the bottom where the tube is entirely occupied by the fluid and is equal to 0 at the top of the tube.  
Although $s(z;t)$ can change continuously from 1 to 0 as $z$ increases, it has been shown \cite{2018_square} that the actual profile of $s(z;t)$ is close to that shown in Fig.~\ref{fig:sketch}(b), 
i.e., $s(z)$ is equal to 1 in the bottom region $0<z<h_0(t)$, and changes to a small value $s^*$ in the narrow transition region of $h_0(t) < z < h_0(t) + \Delta z$, and then decreases gradually to zero in the region $ h_0(t) + \Delta z < z < h_0(t)+h_1(t)$.   
Since the length of the transition region $\Delta z$ is of the order of the tube width $a$ and is much smaller than $h_0(t)$ and $h_1(t)$, we shall ignore this transition region, and assume that $s(z;t)$ changes discontinuously from $1$ to $s^*$ at $z=h_0(t)$ in the following treatment.
Therefore, at a given time $t$, the fluid distribution is described by the bulk
height $h_0(t)$ and the finger profile $s(z,t)$ with $h_0<z<h_0+h_1$.
In the following, we will derive the equations governing the time evolution of
$h_0(t)$ and $s(z;t)$ using Onsager variational principle \cite{DoiSoft, Doi2021}.

To use the Onsager principle, we shall first consider how the free energy of the system
is written as a function of the state variables $h_0(t)$,  $h_1(t)$  and $s(z;t)$.
The total energy is given by $A=A_g + A_{\gamma}$, where $A_g$ and $A_{\gamma}$
stand for the gravitational energy and the interfacial energy, respectively.

\paragraph*{Gravitational energy.}
The gravitational energy is given by
\begin{equation}
\label{eq1}
  A_g = \frac{1}{2} \rho g S h_{\rm 0}^2 + \rho g S \int_{h_{\rm 0}}^{h_{\rm 0}+h_{\rm 1}} s z \, \ud z \, ,
\end{equation}
where $S$ is the cross-section area of the tube.
The change rate of the gravitational energy is
\begin{equation}
  \label{eq:Ag_dot1}
  \frac{\dot{A_g}}{\rho g S} = h_{\rm 0} \dot{h}_{\rm 0}
  + \int_{h_{\rm 0}}^{h_{\rm 0}+h_{\rm 1}} \left( z\frac{\partial s}{\partial t} \right) \ud z
  - (s^* h_{\rm 0})\dot{h}_{\rm 0} \, .
\end{equation}
Here the partial time derivative is denoted by a dot. 
We also used that $s(h_0;t) = s^*$ and $s(h_0+h_1;t)=0$ on the two ends of the finger.

The conservation equation in the finger is given by
\begin{equation}
\label{eq:cons}
  \frac{\partial s}{\partial t} = - \frac{\partial {j_1}}{\partial z} \, ,
\end{equation}
where $j_1(z,t)$ is the volume flux divided by the tube cross-section.
Using the conservation equation and integration-by-part, Eq.~(\ref{eq:Ag_dot1}) becomes
\begin{equation}
  \label{eq:Ag_dot2}
  \frac{\dot{A}_g}{\rho g S} = h_{\rm 0} \dot{h}_{\rm 0}
  + h_0 j_1^*
  + \int_{h_{\rm 0}}^{h_{\rm 0}+h_{\rm 1}} j_1\, \ud z
  - ( s^* h_{\rm 0})\dot{h}_{\rm 0} \, ,
\end{equation}
where $j_1^*$ is the flux at the start of the finger ($z=h_0$).

\paragraph*{Interfacial energy.}
We use $f(s)$ to denote the interfacial energy per length of the system,
which is a function of the local saturation $s$ \cite{2018_square, 2021_rect_sstar}.
The interfacial energy is given by
\begin{equation}
\label{eq2}
  A_{\gamma} = f(1) h_{\rm 0} + \int_{h_{\rm 0}}^{h_{\rm 0}+h_{\rm 1}} f(s) \ud z \, ,
\end{equation}
where the two terms account for the contributions from the bulk and the finger, respectively.
\change{Here we have ignored the energy associated with the transition region between the bulk part and the finger part since it is given by an integral in a small region (much smaller than $h_0$ and $h_1$), and does not affect the motion of $h_0(t)$ and $h_1(t)$.}
  
The change rate of the interfacial energy is
\begin{eqnarray}
  \dot{A}_{\gamma} &=& f(1) \dot{h}_{\rm 0}
    + \int_{h_{\rm 0}}^{h_{\rm 0}+h_{\rm 1}} \left( f'(s) \frac{\partial s}{\partial t} \right) \ud z
    - f(s^*) \dot{h}_{\rm 0} \nonumber \\
  \label{eq:Agamma_dot2}
  &=& f(1) \dot{h}_{\rm 0} + f'(s^*) j_1^*
    + \int_{h_{\rm 0}}^{h_{\rm 0}+h_{\rm 1}} f''(s) \frac{\partial s}{\partial z} j_1 \ud z
    - f(s^*) \dot{h}_{\rm 0} \, .
\end{eqnarray}
Here we define $f'(s) \equiv \ud f(s) / \ud s$ and $f''(s) \equiv \ud^2 f(s) / \ud s^2$.
Again the conservation equation (\ref{eq:cons}) and integration-by-part are used to obtain the second line.

\paragraph*{Change rate of the total energy.}
The change rate of the total energy is then given by Eqs. (\ref{eq:Ag_dot2}) and (\ref{eq:Agamma_dot2})
\begin{eqnarray}
  \dot{A} &=& \rho g S h_{\rm 0} \dot{h}_{\rm 0} + \rho g S \Big[ h_0 j_1^* + \int_{h_{\rm 0}}^{h_{\rm 0}+h_{\rm 1}} j_1 \ud z - ( s^* h_{\rm 0})\dot{h}_{\rm 0} \Big] \nonumber  \\
  &+& f(1) \dot{h}_{\rm 0} + f'(s^*) j_1^* + \int_{h_{\rm 0}}^{h_{\rm 0}+h_{\rm 1}} f''(s) \frac{\partial s}{\partial z} j_1 \ud z - f(s^*) \dot{h}_{\rm 0} \, .
\label{eq:Adot}
\end{eqnarray}

\paragraph*{Dissipation function.}
The dissipation function $\Phi$ is defined as the half of the energy dissipation rate when the system is evolving at a given rate of the state variables.  
In the present problem, we consider the situation that the profile is changing at rate $\dot h_0$, $\dot h_1$, and $\partial s/\partial t$, which is represented by the flux $j(z;t)$.
\change{Using the lubrication approximation, the dissipation function per unit length is
written as a quadratic function of the local flux, $\frac{1}{2} \zeta(s) j^2(z;t)$.
The friction constant $\zeta(s)$ is a function of the local saturation
$s(z;t)$ under lubrication approximation \cite{Ransohoff1988, 2018_square}.}
The total dissipation function is given by integrations over the bulk and the finger
\begin{eqnarray}
  \Phi &=& \frac{1}{2} \zeta(1) j_0^2 h_0
           + \frac{1}{2} \int_{h_{\rm 0}}^{h_{\rm 0}+h_{\rm 1}} \zeta(s) j_1^2 \ud z \nonumber \\
  &=& \frac{1}{2} \zeta(1) \left[ j_1^* + (1-s^*)\dot{h}_0 \right]^2 h_0
           + \frac{1}{2} \int_{h_{\rm 0}}^{h_{\rm 0}+h_{\rm 1}} \zeta(s) j_1^2 \ud z \, .
\label{eq:Phi}
\end{eqnarray}
Here $j_0$ is the flux in the bulk region.
In the second line we have used the conservation relation $j_0 = j_1^* + (1-s^*)\dot{h}_{\rm 0}$.

\paragraph*{Time evolution equations.}
The Rayleighian of the system is $\mathscr{R} = \dot{A} + \Phi$, where the change rate of
the energy and the dissipation function are given by Eqs. (\ref{eq:Adot}) and (\ref{eq:Phi}), respectively.
The time-dependent variables are the flux in the finger $j_1$, the flux at the entrance
of the finger $j_1^*$, and the bulk velocity $\dot{h}_0$.  The time evolution equations then
can be obtained using Onsager variational principle \cite{DoiSoft}.

The variation of $\mathscr{R}$ with $j_1$ is
\begin{eqnarray}
  && \rho g S + f''(s) \frac{\partial s}{\partial z} + \zeta(s) j_1 = 0  \\
  &\Rightarrow \quad & j_1 = -\frac{f''(s)}{\zeta(s)} \frac{\partial s}{\partial z} - \frac{\rho g S}{\zeta(s)} \, .
\label{eq:j1}
\end{eqnarray}
Combining with the conservation equation (\ref{eq:cons}), we obtain a
diffusion-like partial differential equation (PDE) governing the finger dynamics
\begin{equation}
\label{eq:pde0}
  \frac{\partial s}{\partial t} = \frac{\partial }{\partial z} \Big[ D(s) \frac{\partial s}{\partial z}+\frac{\rho g S}{\zeta(s)} \Big],
\end{equation}
where $D(s) \equiv f''(s)/\zeta(s)$ is the diffusion constant.

The boundary conditions are $s(h_0) = s^*$ and $s(h_0+h_1) = 0$.
They denote the conditions at moving boundaries $h_0(t)$ and $h_1(t)$.
We perform a change of variables by $z' = z - h_0$, $\tau = t$, then the PDE (\ref{eq:pde0}) becomes
\begin{equation}
\label{eq:pde}
  \frac{\partial s}{\partial \tau} = \frac{\partial }{\partial z'} \Big[ D(s) \frac{\partial s}{\partial z'}+\frac{\rho g S}{\zeta(s)} \Big] + \dot{h}_{\rm 0}\frac{\partial s}{\partial z'},
\end{equation}
with the fixed boundary conditions $s(z'= 0) = s^*$ and $s(z'=h_1) = 0$.

The variations of $\mathscr{R}$ with $j_1^*$ and $\dot{h}_0$ are
\begin{eqnarray}
  \label{eq:dRdj1s}
  && \rho g S h_0 + f'(s^*) + \zeta(1)h_0 \left[ j_1^* + (1-s^*)\dot{h}_0 \right] = 0, \\
  \label{eq:dRdh0d}
  && \rho g S h_0 (1-s^*) + f(1) - f(s^*) + \zeta(1)h_0 \left[ j_1^* + (1-s^*)\dot{h}_0 \right] (1-s^*) = 0.
\end{eqnarray}
Equations (\ref{eq:dRdj1s}) and (\ref{eq:dRdh0d}) lead to the following expression for $s^*$
\begin{equation}
\label{eq:sstar}
  \frac{f(1) - f(s^*)}{1-s^*} = f'(s^*).
\end{equation}
This expression (\ref{eq:sstar}) has the same form as in the systems where the gravitational effect is absent \cite{2021_rect_sstar, 2018_square}, indicating that the gravity has no effect on $s^*$.

Combining Eqs.~(\ref{eq:j1}) and (\ref{eq:dRdj1s}), we obtain an ordinary differential
equation (ODE) for $h_0$
\begin{equation}
\label{eq:ode}
  h_0 \left[ \left( - D(s^*)\frac{\partial s}{\partial z'} \bigg | _{z'=0}
      - \frac{\rho g S}{\zeta(s^*)} \right) + (1-s^*) \dot{h}_0 \right]
  = - \frac{1}{\zeta(1)} \big( f'(s^*) + \rho g S h_0 \big).
\end{equation}

To summarize, the time evolution of the bulk and the finger are given by a pair of coupled
differential equations.
The bulk dynamics is governed by the ODE (\ref{eq:ode}), but the information at the finger
entrance ($\partial s/\partial z' |_{z'=0}$) is needed.
The finger dynamics is given by the PDE (\ref{eq:pde}), where the bulk velocity $\dot{h}_0$ is required.

%%%%%%%%%%%%%%%%%%%%%%%%%%%%%%%%%%%%%%%%%%%%%%%%%%%%%%%%%%%%%%
\section{Results and Discussions}
\subsection{Bulk-only solution}

Before analyzing the above set of equations, we first consider the hypothetical case
that there are no fingers, and derive the time evolution equation for the bulk flow only.
The Rayleighian for the bulk-only situation is
\begin{equation}
  \mathscr{R} = \rho g S h_0 \dot{h}_0 + f(1) \dot{h}_0 + \frac{1}{2} \zeta(1) h_0 \dot{h}_0 ^2 \, .
\end{equation}
The dynamics is given by $\delta \mathscr{R} / \delta {\dot{h}_0} = 0$, which leads to
\begin{equation}
  \label{eq:h0dot_J}
  \dot{h}_0 = - \frac{f(1)}{\zeta (1)} \frac{1}{h_0} - \frac{\rho g S}{\zeta(1)} \, .
\end{equation}

The equilibrium height is given by $\dot{h}_0 = 0$,
\begin{equation}
  \label{eq:HJ}
  H_{\rm J} =  (h_0)_{\rm eq} = - \frac{f (1)}{\rho g S}.
\end{equation}
We shall call this height $H_{\rm J}$ Jurin's height \cite{Jurin1718}.
For a circular tube of radius $r$, $f(1)=-2\pi r \gamma \cos\theta$ 
%($\gamma$ and $\theta$ being the surface tension and the equilibrium contact angle of the fluid) 
and $S=\pi r^2$, we get the well-known Jurin's height $H_{\rm J} = 2\gamma \cos\theta / (\rho g r)$.
For a square tube of side length $2a$, $f(1)=-8a \gamma \cos\theta$ and $S=4a^2$, the Jurin's height is $H_{\rm J} = 2 \gamma\cos\theta / (\rho g a)$.

We can make the evolution equations dimensionless by scaling the length with Jurin's height $H_{\rm J}$ and the time with
\begin{equation}
  T_{\rm J} = \frac{H_{\rm J}^2}{|f(1)|} \zeta(1) = \frac{ |f(1)| \zeta(1) }{ (\rho g S)^2 } \, .
\end{equation}
Equation (\ref{eq:h0dot_J}) then becomes
\begin{equation}
  \frac{\ud \tilde{h}_0}{\ud \tilde{t}} = \frac{1}{\tilde{h}_0} - 1 \, .
\end{equation}
Here the symbols with tilde represent the dimensionless variables.

The solution to the above equation is
\begin{equation}
  \label{eq:h0J}
  \tilde{h}_0 + \ln \big( 1 - \tilde{h}_0 \big) = - \tilde{t} .
\end{equation}

\paragraph*{Short-time limit.}
In the limit of $\tilde{t} \rightarrow 0 $ and $\tilde{h}_0 \ll 1 $, we have
\begin{equation}
  \ln ( 1 - \tilde{h}_0 ) \simeq - \tilde{h}_0 - \frac{1}{2} \tilde{h}_0 ^2 - \frac{1}{3} \tilde{h}_0 ^3 - \cdots
\end{equation}
Keeping only the first two terms, Eq.~(\ref{eq:h0J}) becomes
\begin{equation}
  \tilde{h}_0 = \sqrt{ 2 \tilde{t}}.
\end{equation}
This is the $t^{1/2}$ scaling of Lucas-Washburn \cite{Lucas1918, Washburn1921}.

\paragraph*{Long-time limit.}
In the limit of $\tilde{t} \rightarrow \infty $, the magnitude of the first term in Eq. (\ref{eq:h0J}) is much less than that of the second term and can be neglected.
We then get
\begin{equation}
   \ln \big( 1 - \tilde{h}_0 \big) = - \tilde{t} \quad
   \Rightarrow \quad \tilde{h}_0 = 1 - e ^ {-\tilde{t}} \, .
\end{equation}
The bulk height approaches its asymptotic value of $H_{\rm J}$ in a manner of $e^{-\tilde{t}}$.

\subsection{Equilibrium states}

We now come back to the coupled equations that include the effect of the finger flow. 
We first discuss the equilibrium states attained in the limit of very long time.

%=========================
\paragraph*{Equilibrium bulk height.}
At equilibrium, all fluxes are zero.
The equilibrium height of the bulk is given by Eq. (\ref{eq:ode}) where all the flux terms on the left-hand-side are zero,
\begin{equation}
  (h_0)_{\rm eq} = - \frac{ f'(s^*) }{ \rho g S } \, .
\end{equation}
Compared to the Jurin's height (\ref{eq:HJ}), we can see
\begin{equation}
  \label{eq:bulk_h0}
  (\tilde{h}_0)_{\rm eq} = \frac{ (h_0)_{\rm eq} }{ H_{\rm J} } = \frac{ f'(s^*) }{ f(1) } < 1 \, .
\end{equation}

\begin{figure}[htbp]
  \centering
  \includegraphics[width=0.5\columnwidth]{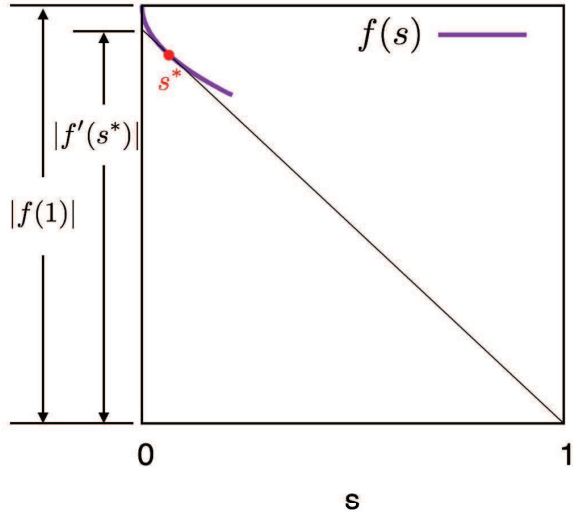}
  \caption{The free energy curve $f(s)$ and the determination of $s^*$.}
  \label{fig:freeE}
\end{figure}

The inequality can be explained by examining the free energy curve $f(s)$ in Fig.~\ref{fig:freeE}.
At small saturation, the free energy is a decreasing and convex function of $s$.
Geometrically, the condition (\ref{eq:sstar}) for $s^*$ corresponds to drawing a straight line passing through two points $(1, f(1))$ and $(s^*, f(s^*))$, and this line is also tangential to $f(s)$ curve.
The slope of this line is $f'(s^*)$, thus it intersects with the $y$-axis at a point whose height is $f'(s^*)$.
Due to the convexity of the $f(s)$ curve, we always have $|f'(s^*)| < |f(1)|$.
From Eq. (\ref{eq:bulk_h0}), this indicates that the equilibrium bulk height is always less than the Jurin's height when the fingers are present. % (ignoring the fingers).

For a square tube with side length $2a$, assuming the contact angle between the liquid and the tube is $\theta$, we have the free energy density of both the bulk and the finger \cite{2018_square}
\begin{eqnarray}
  \label{eq34}
  f(1) &=& -A_0 a \gamma, \quad A_0 = 8 \cos \theta \\
  \label{eq35}
  f(s) &=& - A_1 \sqrt{s} a \gamma, \quad  A_1 = 8\sqrt{ \cos ^2 \theta - \sin \theta \cos \theta - (\frac{\pi}{4} - \theta)}
\end{eqnarray}

Combining Eqs. (\ref{eq:sstar}), (\ref{eq34}) and (\ref{eq35}), we can get
\begin{equation}
\label{eq:star_square}
  s^* = \frac{\cos \theta - \sqrt{ \sin \theta \cos \theta + ( \frac{\pi}{4} - \theta )}}{\cos \theta + \sqrt{ \sin \theta \cos \theta + ( \frac{\pi}{4} - \theta )}}.
\end{equation}

\begin{figure}[htbp]
  \centering
  \includegraphics[width=0.8\columnwidth]{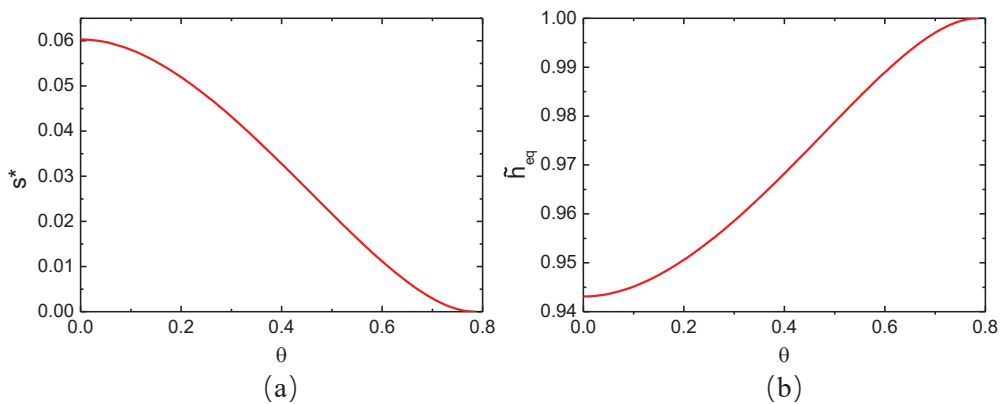}
  \caption{ (a) The equilibrium saturation $s^*$ is plotted against the contact angle $\theta$ ($0 \le \theta \le \pi/4$). (b) The equilibrium bulk height $\tilde{h}_{\rm eq}$ is plotted against the contact angle.}
  \label{fig:bulk_eq}
\end{figure}

When $\theta = 0$, $s^*$ is about 0.06, which agrees with the previous result of Ref. \cite{2018_square}.
We show the change of $s^*$ with $\theta$ in Fig. \ref{fig:bulk_eq}(a).
As the contact angle $\theta$ increases, $s^*$ decreases.
When $\theta=\pi/4$, $s^* = 0$.
The fingers vanish for contact angle $\theta > \pi/4$ in a square tube.
This also agrees with the Concus-Finn's condition \cite{Concus1969} for the existence of unbounded surface in a corner.

The equilibrium bulk height as a function of the contact angle is given by Eqs. (\ref{eq:bulk_h0}) and (\ref{eq:star_square}),
\begin{equation}
  \tilde{h}_{\rm eq} = \frac{1}{2} \frac{ \cos\theta + \sqrt{\sin\theta \cos\theta + (\frac{\pi}{4} - \theta) } }{\cos\theta} \, .
\end{equation}
The result is shown in Fig. \ref{fig:bulk_eq}(b).
For $\theta=0$, we have $h_{\rm eq} = \frac{1}{4}(2+\sqrt{\pi}) H_{\rm J}$. 
This agrees with the result in Ref. \cite{Bico2002} where experiments were also conducted to validate this result. 
The bulk height is inversely related to $s^*$: bigger $s^*$ is, smaller height the bulk can reach. 
For square tubes, the effect of the finger on the equilibrium bulk height is rather small (only about 5\% reduction with respect to the Jurin's height) due to the smallness of $s^*$ (about 0.06).
The effect might be more prominent in tubes with bigger $s^*$ \cite{Keita2016, 2021_rect_sstar, 2021_rect_LW}.

%=====================================
\paragraph*{Equilibrium finger profile.}
Letting $j_1=0$ in Eq. (\ref{eq:j1}), we obtain the equilibrium profile for the finger
\begin{equation}
  \frac{\partial s}{\partial z} = - \frac{\rho g S}{f''(s)} \, .
\end{equation}

For square tubes, we have $f''(s)=\frac{1}{4} A_1 s^{-3/2} a \gamma$.
This leads to the following equation
\begin{equation}
  \frac{\partial s}{\partial z} = - \frac{16\rho g a}{A_1 \gamma} s^{3/2}.
\end{equation}
The dimensionless form is
\begin{equation}
  \frac{\partial s}{\partial \tilde{z}} = - \frac{4 A_0}{A_1} s^{3/2}.
\end{equation}
The solution to the above equation is
\begin{equation}
  \label{eq:finger_eq}
  \frac{1}{\sqrt{s_{\rm eq}}} - \frac{1}{\sqrt{s^*}} = \frac{2A_0}{A_1} \tilde{z}', \quad {\rm or} \quad
  s_{\rm eq}(\tilde{z}') = \Bigg( \frac{2 A_0}{A_1} \tilde{z'} + {s^*} ^{- \frac{1}{2}} \Bigg)^{-2}.
\end{equation}

\begin{figure}[htbp]
  \centering
    \includegraphics[width=0.6\columnwidth]{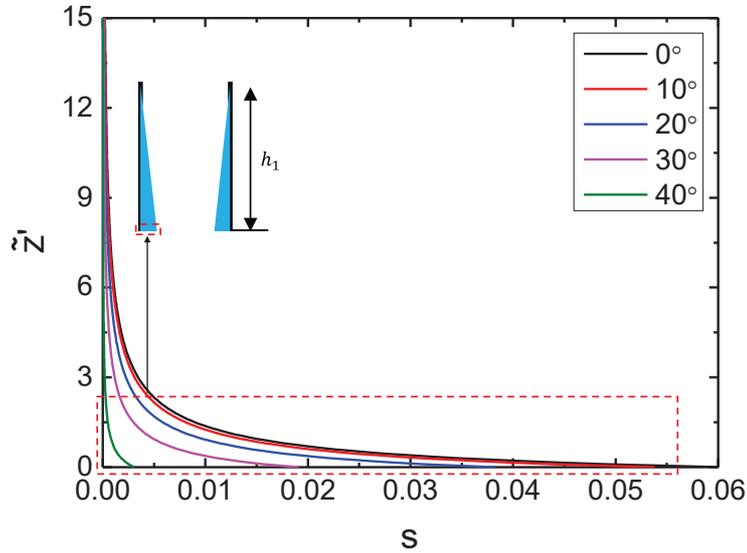}
    \caption{The equilibrium profiles of the finger for different contact angle. The curves represent the relation between the saturation $s$ and the height $\tilde{z}'$.}
  \label{fig:finger_eq}
\end{figure}

The equilibrium profiles of the finger for different contact angle are shown in Fig. \ref{fig:finger_eq}. 
We can visually get the fingers become thinner as the contact angle increases, the results are consistent with Ref. \cite{Gurumurthy2018a}.

%%%%%%%%%%%%%%%%%%%%%%%%%%%%%%%%%%%%%%%%%%%%%%%%%
\subsection{Dynamics of Capillary Rising}

We now proceed to analyze the dynamics of the coupling between the bulk flow and the finger flow.
Here we focus on the fully-wetted case ($\theta = 0$), for which the friction coefficient $\zeta(s)$ has been calculated in Ref. \cite{2018_square}.
For square tube, the parameters needed for the calculation are given as follows
\begin{eqnarray}
  s^*  &=& 0.0603178 \\
  S    &=& 4 a^2    \\
  f(1) &=& -A_0 a \gamma, \quad A_0 = 8 \\
  f(s) &=& -A_1 s^{1/2} a \gamma, \quad A_1 = 8 \sqrt{1-\frac{\pi}{4}} \\
  \zeta(1) &=& B_0 \eta, \quad  B_0 = 28.46 \\ %28.4558 \\
  \zeta(s) &=& B_1 s^{-2} \eta, \quad B_1 = 78.50 \\ %78.5031 \\
  D(s) &=& \frac{f''(s)}{\zeta(s)} = \frac{A_1}{4 B_1} s^{1/2} \frac{a \gamma}{\eta}
\end{eqnarray}
\change{The detailed derivations of $f(s)$ and $\zeta(s)$ are given in the Supporting Information.}
Here we have made a notation change.
Previously in Ref. \cite{2018_square}, we wrote the dissipation function in term of the volume flux $Q(z)$, and the corresponding friction coefficient is $\xi(s)$, i.e., $\Phi = \frac{1}{2} \int \ud z \, \xi(s) [Q(z)]^2$.
In the current work for general tubes, it is more convenient to express the dissipation function in term of the volume flux divided by the cross-section area, $j(z)=Q(z)/S$.
This leads to a different friction coefficient $\Phi = \frac{1}{2} \int \ud z \, \zeta(s) [j(z)]^2 = \frac{1}{2} \int \ud z \, \xi(s) [Q(z)]^2$.
These two friction coefficients are related by $\zeta(s) = \xi(s) S^2$.

The coupled time-evolution equations for square tubes are
\begin{eqnarray}
\label{eq52}
  \frac{\partial s}{\partial \tau} &=& \frac{\partial }{\partial z'}
  \Bigg( \frac{A_1}{4 B_1} \frac{a \gamma}{\eta}
  s^{1/2} \frac{\partial s}{\partial z'}
  + \frac{4}{B_1} \frac{\rho g a^2}{\eta} s^2 \Bigg)
  + \dot{h}_0 \frac{\partial s}{\partial z'},  \\
\label{eq53}
  \frac{\mathrm{d} h_0}{\mathrm{d} \tau} &=& \frac{1}{(1-s^*)} \frac{ a \gamma}{\eta} \Bigg( - \frac{4}{B_0} \bigg( \frac{- A_1 }{8 \sqrt{s^*}} + \frac{a \rho g}{\gamma} h_0 \bigg) \frac{1}{h_0} + \frac{A_1}{4 B_1} \sqrt{s^*} \frac{\partial s}{\partial z'} \bigg | _{z' = 0} + \frac{4 a \rho g (s^*) ^2}{\gamma B_1} \Bigg).
\end{eqnarray}

Scaling the length with $H_{\rm J}$ and the time with $T_{\rm J}$, we obtain the dimensionless form
\begin{eqnarray}
\label{eq54}
  \frac{\partial s}{\partial \tilde{\tau}} &=& \frac{\partial }{\partial \tilde{z'}} \Bigg( \frac{A_1 B_0}{4 A_0 B_1} s^{1/2} \frac{\partial s}{\partial \tilde{z'}} + \frac{B_0}{ B_1} s^2 \Bigg) + \dot{\tilde{h}}_0 \frac{\partial s}{\partial \tilde{z'}},  \\
\label{eq55}
  \frac{\mathrm{d} \tilde{h}_0}{\mathrm{d} \tilde{\tau}} &=& \frac{1}{(1-s^*)} \Bigg( \bigg( \frac{ A_1 }{2 A_0 \sqrt{s^*}} - \tilde{h}_0 \bigg) \frac{1}{\tilde{h}_0} + \frac{A_1 B_0}{4 A_0 B_1} \sqrt{s^*} \frac{\partial s}{\partial \tilde{z'}} \bigg | _{\tilde{z'} = 0} + \frac{B_0}{B_1} ( s^* ) ^2 \Bigg).
\end{eqnarray}
%The units of $H_J$ and $T _J$ are $[m]$ and $[s]$, respectively.

A cautionary note is that even though the side length of the tube $2a$ does not appear in Eqs. (\ref{eq54}) and (\ref{eq55}), it does not imply that the capillary rise can take place in tubes of arbitrary width.
The capillary rise occurs in the situation that the tube width is small in comparison with the capillary length $\sqrt{{\gamma}/{\rho g}}$ \cite{dBQ}.
This can be expressed as ${\gamma}/{\rho g} \gg a^2$, which is equivalent to the condition that the tube width $2a$ is much smaller than the Jurin's height $H_{\rm J} \simeq \gamma/\rho g a$.

%In this section, we show the dynamic evolution process for the bulk flow $\tilde{h}_0$ and the finger flow $\tilde{h}_1$, and $\tilde{h}_1$ is actually obtained by the diffusion of saturation parameter $s$.

\begin{figure}[htbp]
  \centering
    \includegraphics[width=1\columnwidth]{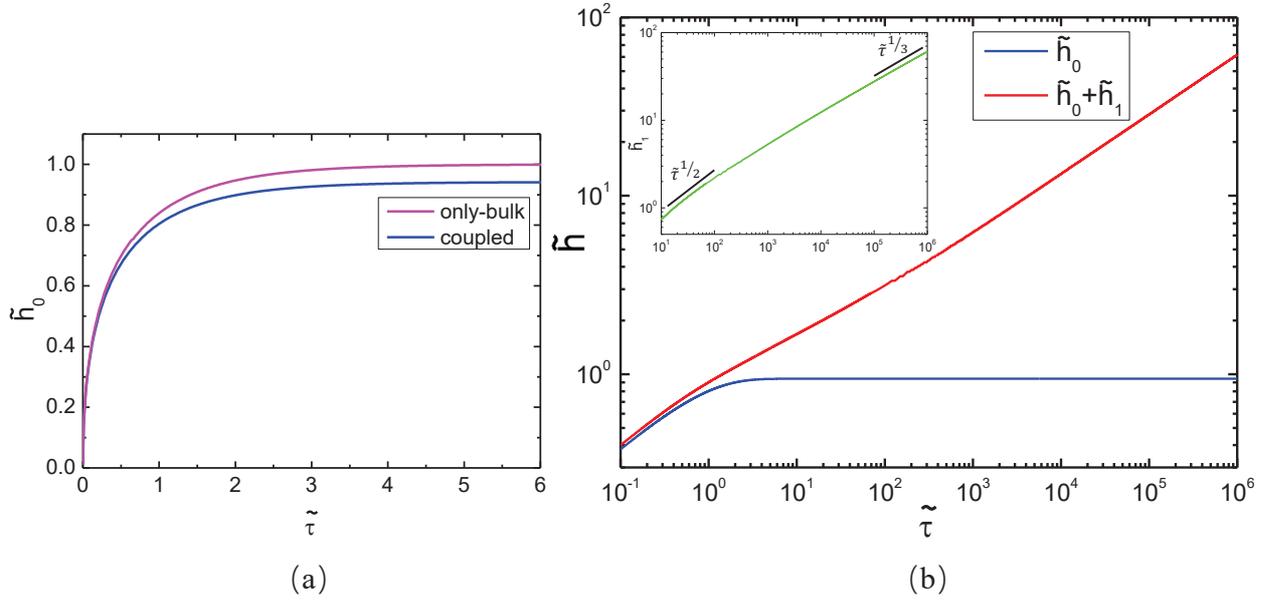}
    \caption{(a) The time evolution of $\tilde{h}_0$. (b) The time evolution of $\tilde{h}_0 + \tilde{h}_1$. The inset shows the finger flow evolves with time.}
  \label{fig:h01}
\end{figure}

We solved the coupled PDE (\ref{eq54}) and ODE (\ref{eq55}) using the Euler forward difference method.
Figure \ref{fig:h01}(a) shows the time evolution of the bulk height.
The blue solid line represents the numerical solutions and the purple solid line denotes the bulk-only case [Eq. (\ref{eq:h0J})].
The bulk height $\tilde{h}_0$ rises quickly with time, reaching an equilibrium value at near $\tilde{\tau} \simeq 4$. The equilibrium value of the numerical solution for the bulk-only and the coupled case at long-time regime gives about $1$ and $0.943$, respectively.

Figure \ref{fig:h01}(b) shows both the bulk height $\tilde{h}_0$ and the finger tip height $\tilde{h}_0 + \tilde{h}_1$. 
%Here we should state that even $\tilde{h}_0$ visually stays at an equilibrium height, in our numerical calculation results it still keeps rising up at a extremely small value $ \delta \tilde{h}_0 = 10^{-5}$ at $\tilde{\tau} = 30$, and with time goes on, the increment becomes smaller. 
Although the bulk height reaches a plateau in a short time, the finger keeps rising with time. 
The inset in Fig. \ref{fig:h01}(b) shows only the time evolution of the finger length $\tilde{h}_1$. 
The fitting of numerical calculations indicates that $\tilde h_1$ increases as 0.2328 $\tilde \tau^{1/2}$ and 0.5903 $\tilde \tau^{1/3}$ in the short and long lime regions respectively.
%takes a time scaling law transition from $\tilde{\tau} ^{1/2}$ to $\tilde{\tau} ^{1/3}$, and the prefactors are 0.2328 and 0.5903, respectively. 
The result indicates that the finger flow obeys the Lucas-Washburn scaling law in relatively short time where the effect of gravity is negligible. 
In a long time region, the gravity becomes more important, which decreases the speed of the finger flow, changing the scaling law from $t^{1/2}$ to $t^{1/3}$.

\begin{figure}[htbp]
  \centering
    \includegraphics[width=0.8\columnwidth]{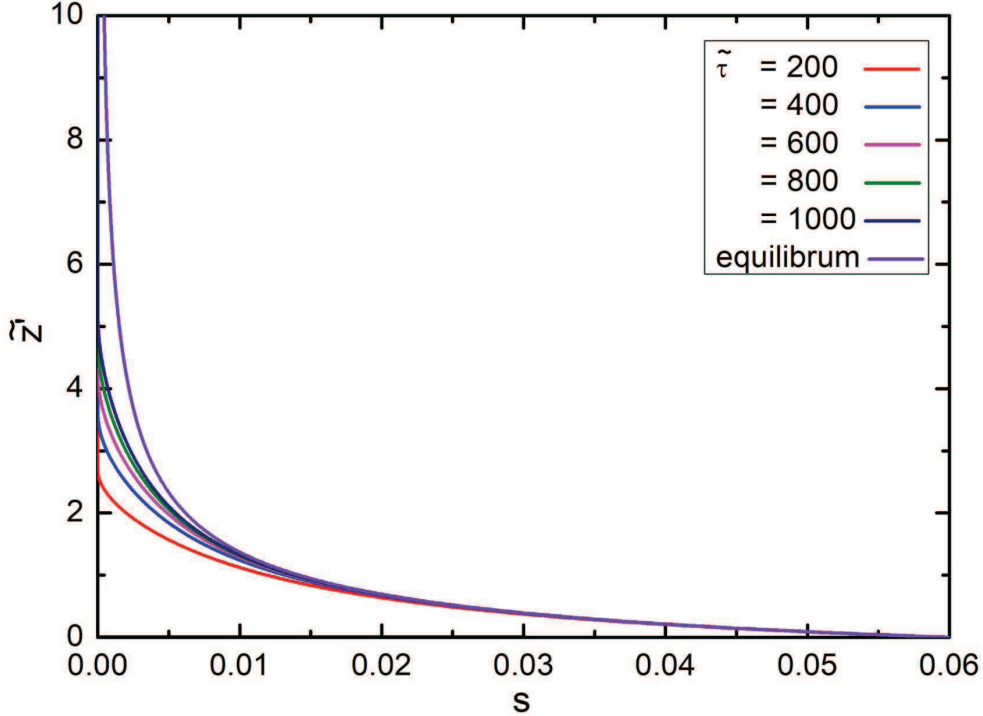}
    \caption{The finger profile at different times ($\tilde{\tau} = 200, 400, 600, 800, 1000$ from bottom to the top) and the equilibrium state.}
  \label{fig:finger_t}
\end{figure}

Figure \ref{fig:finger_t} shows the finger profiles at different times.
The height of $\tilde{h}_1$ is obtained from the position of the tip in $\tilde{z}'$ axis. 
As time goes by, the finger imbibes along the $\tilde{z}'$ axis, and eventually approaches the equilibrium profile [Eq. (\ref{eq:finger_eq})].

%%%%%%%%%%%%%%%%%%%%%%%%%%%%%%%%%%%%%%
\subsection{Comparison with experiments}

\begin{figure}[htbp]
  \centering
    \includegraphics[width=0.9\columnwidth]{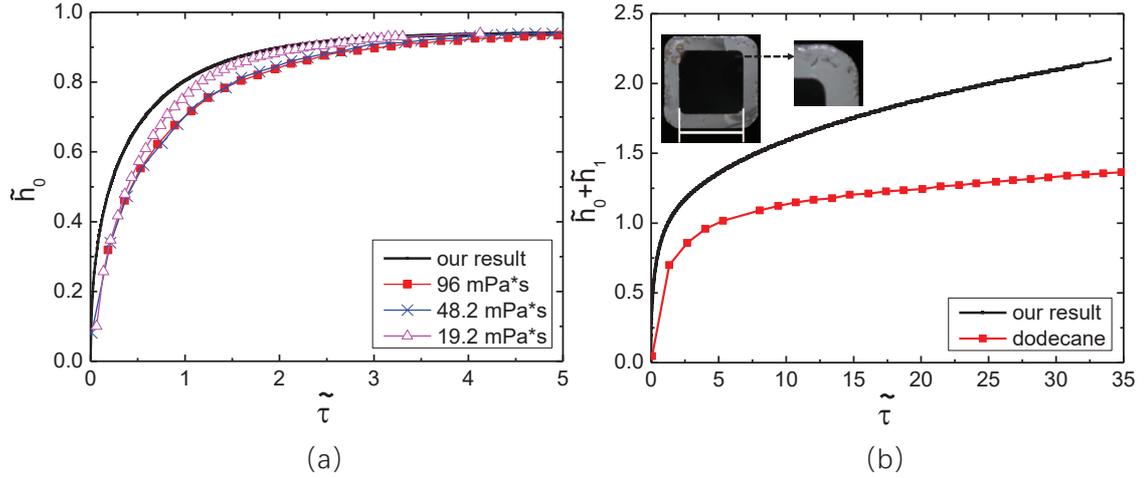}
    \caption{The comparison between our results and experimental measurement. 
(a) The dimensionless height of the bulk part $\tilde h_0$ is plotted against the dimensionless time $\tilde \tau$. The solid black curve represents the numerical solution of our model. 
%The fuchsin inverted triangle line, the blue triangle line, and the red solid dot line represent PDMS oils of different viscosities $ \eta = $ 96, 48.2 and 19.2 mPa$\cdot$s ($ \pm $5$\%$ ), the corresponding densities of which are 960, 950 and 930 kg$\cdot$m$^{-3}$. 
The experimental results of PDMS for viscosities  $\eta=$ 96, 48.2 and 19.2 mPa$\cdot$s are shown by the curves with symbols. 
%The surface tension of these oils is constant at 19.8 mN$\cdot$m$^{-1}$, the contact angle is 0${^\circ}$, and the tube length used is 0.6$mm$. 
To calculate the reduced height and the reduced time, we used the following values (density $\rho =$ 960, 950, and 930 kg/m$^3$, surface tension $\gamma=$ 19.8 mN/m, equilibrium contact angle $\theta=0$, and tube side length $2a= $ \change{0.6} mm). Experimental results are from Ref. \cite{Ouali2013}. 
(b) The reduced total height $\tilde h_0 + \tilde h_1$ is plotted against the reduced time $\tilde \tau$.
The solid black curve represents the numerical solution of our theory, and the red square line represents the experimental results of Dodecane liquid with surface tension $\gamma=$ 24  mN/m, viscosity $\eta=$ 1.34 mPa$\cdot$s, density $\rho=$ 750 kg/m$^{-3}$, equilibrium contact angle $\theta=$ 0, and tube side length $2a=$ 0.5 mm. Experimental results are from Ref. \cite{Wijnhorst2020}. 
Image of the square tube in the inset is adapted with permission from Ref. \cite{Wijnhorst2020}. Copyright 2020 American Chemical Society.}
  \label{fig:compare}
\end{figure}

We compare our solutions with reported experimental results, for fluids with zero contact angles.
% in Ref. \cite{Ouali2013} and Ref. \cite{Wijnhorst2020}. 
For the bulk flow $\tilde{h}_0(t)$, Ouali \emph{et al.} measured the capillary rising in square tubes using PDMS liquids of various viscosities \cite{Ouali2013}.
Wijnhorst \emph{et al.} reported the capillary rising for the tips of the fingers \cite{Wijnhorst2020}. 
The values of parameters are shown in the caption in Fig. \ref{fig:compare}. 
While the data graphs in the original references are with dimension, here we show the results in a dimensionless form using $H_{\rm J}$ and $T_{\rm J}$. 
%As for the results in Ref. \cite{Ouali2013}, taking the experimental parameters in we can get $H_J = $ 0.01403$m$, 0.01417$m$, 0.01448$m$ and $T_J = $ 11.3157$s$, 5.7954$s$, 2.4106$s$ for $ \eta = $ 96, 48.2 and 19.2 mPa$\cdot$s, respectively. For Ref. \cite{Wijnhorst2020} we get $H_J = $ 0.02612$m$ and $T_J = $ 0.299$s$.

Figure \ref{fig:compare}(a) shows the comparison of the bulk dynamics \cite{Ouali2013}.
Here the parameters are $H_{\rm J} = $ 0.01403 m, 0.01417 m, 0.01448 m and $T_{\rm J} = $ 11.3157 s, 5.7954 s, 2.4106 s for $\eta =$ 96, 48.2 and 19.2 mPa$\cdot$s, respectively. 
There is a reasonable agreement between our numerical results and the experimental measurement. 
The equilibrium bulk height are all smaller than 1. 
The bulk heights of three PDMS fluid are about 0.0131 m, scaled by the corresponding $H_{\rm J}$ we get the dimensionless bulk heights as 0.9337, 0.9238 and 0.9047, respectively. 
The small differences are difficult to measure in experiments, possibly due to the finite length of the tube.

Figure \ref{fig:compare}(b) shows the comparison of the finger dynamics \cite{Wijnhorst2020}.
Here the tip height $\tilde h_0 +\tilde h_1$ is shown and the agreement is not good.
Though various reasons are conceivable, we think that the main reason for this discrepancy is due to the fact that the corners of the tubes used in the experiments are not very sharp as it is shown by the picture in the inset of Fig. \ref{fig:compare}(b).
%The finger dynamics can be slowed down by the roundness of the corner \cite{Ransohoff1988}. 
\change{If the corner is round, there is less area of liquid-solid contact in comparison to the sharp corner. 
This leads to a reduction of the driving force for the finger rising, and the meniscus rising of the finger eventually stops at some equilibrium height. 
On the other hand, the roundness of the corner introduces extra flow resistance \cite{Ransohoff1988}. 
These effects lead to the slowdown of the finger flow.}
%With intense curiosity, we find that the corners of the square tube used in Ref. \cite{Wijnhorst2020} are not so sharp, which is expressed in the inset of Figure \ref{fig7}(b). Analytically thinking, the exist of arcs in corners of the square tube will decrease the evolutionary speed, that why the theoretical flow is faster than the experimental flow.

\section{Conclusion}

To conclude, we have studied the capillary phenomena in cornered tubes with gravity effect, and derived coupled time-evolution equations for the bulk flow and the finger flow. 
We have shown that at equilibrium the bulk height is reduced from Jurin's height due to the presence of the fingers.  By solving the coupled evolution equations numerically, we have shown that the bulk height reaches an equilibrium value in \change{short} time, while the finger height keeps increasing following the $t^{1/3}$ scaling law.

%the equilibrium state of both the bulk height and the saturation profiles of finger for different contact angles. The bulk height is reduced when the fingers are considered. 
%We also performed the numerical evolutionary processes for the bulk flow and the finger flow. The bulk height reaches a plateau value at short time, while the finger continues rising with a time scaling $t^{1/3}$ for long time.

\begin{acknowledgments}
This work was supported by the National Natural Science Foundation of China (NSFC) through the Grant No.21774004 (to JZ).
\end{acknowledgments}

\bibliography{cite}

% uncomment the following lines if using preprint endfloats
%\listoffigures

\end{document}

\title{Supplementary Information: \\
\emph{Capillary rising in a tube with corners}}

\author{Chen Zhao}
\affiliation{Center of Soft Matter Physics and Its Applications, Beihang University, Beijing 100191, China}
\affiliation{School of Physics, Beihang University, Beijing 100191, China}

\author{Jiajia Zhou}
\email[]{zhouj2@scut.edu.cn}
\affiliation{South China Advanced Institute for Soft Matter Science and Technology, School of Emergent Soft Matter, South China University of Technology, Guangzhou 510640, China}
\affiliation{Guangdong Provincial Key Laboratory of Functional and Intelligent Hybrid Materials and Devices, South China University of Technology, Guangzhou 510640, China}

\author{Masao Doi}
\affiliation{Center of Soft Matter Physics and Its Applications, Beihang University, Beijing 100191, China}
\affiliation{Wenzhou Institute, University of Chinese Academy of Science, 
Wenzhou, Zhejiang 325000, China}
\maketitle

\tableofcontents

\newpage
\section{The derivation of the interfacial energy density for finger part}
\begin{figure}[htbp]
  \centering
    \includegraphics[width=0.5\columnwidth]{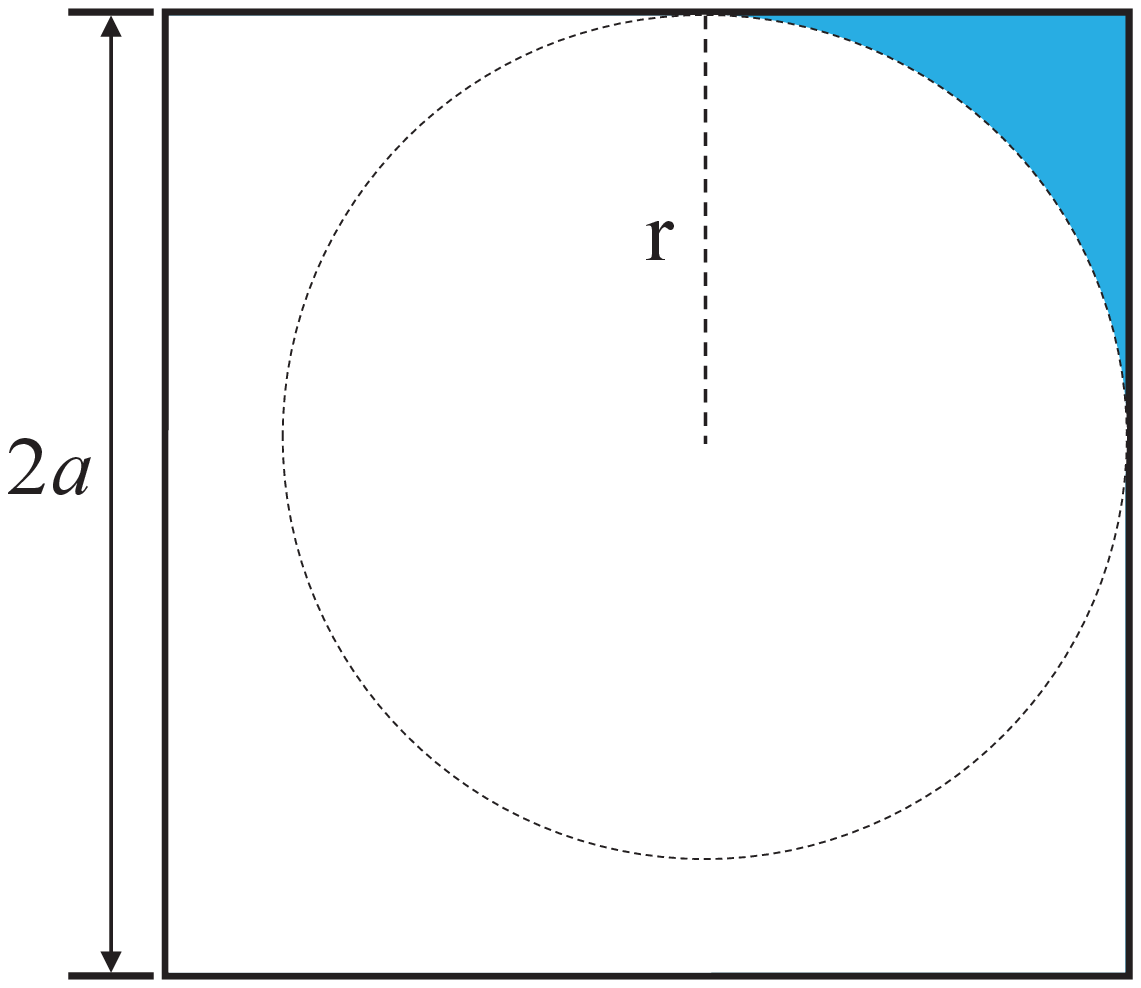}
    \caption{The saturation for finger region as the contact angle $\theta _E = 0$. The tube's side length is $2a$, the radius of the inscribed circle is $r$.}
  \label{fig:square_finger_f_s}
\end{figure}

For the case that the contact angle between the liquid and the solid surface is zero, we have the saturation state of the finger part for any cross-section shown in Fig. \ref{fig:square_finger_f_s}.

The saturation can be calculated as
\begin{equation}
  \label{saturation_1}
  s = \left( 1-\frac{\pi}{4} \right) \frac{r^2}{a^2},
\end{equation}
which leads to the expression of $r$ as a function of the saturation $s$ 
\begin{eqnarray}
\label{r_s}
r = a \sqrt{\frac{1}{(1-\pi/4)}}s^{1/2} .
\end{eqnarray}

The interfacial energy density is
\begin{equation}
  \label{f_s_2}
  f(s) = 8 r (\gamma_{SL} - \gamma_{SV})+2\pi r \gamma  =  -8(1-\pi/4)r\gamma ,
\end{equation}
where Young' equation $ \gamma_{SV} = \gamma_{SL} + \gamma \cos{\theta_E} = \gamma_{SL}$ is used. From Eq. (\ref{r_s}) we get
\begin{eqnarray}
\label{f_s_3}
f(s) = -8 \sqrt{(1-\pi/4)} s^{1/2} a\gamma .
\end{eqnarray}

\newpage
%%%%%%%%%%%%%%%%%%%%%%%%%%
\section{The derivation of the friction coefficient for finger part.}

For the capillary dynamics, we assume the fluid velocity is mostly along the tube axis, $\mathbf{v} \simeq (0, 0, v_z)$ where the $x$- and $y$-components are nearly zero.
The fluid flow satisfies the Stokes equation
\begin{equation}
   \label{eq:Stokes}
   \eta \Big{(} \frac{\partial ^2 v_z}{\partial x^2} + \frac{\partial ^2 v_z}{\partial y^2} \Big{)} = \frac{\partial p}{\partial z},
\end{equation}
where $\partial p / \partial z$ is the pressure gradient in $z$ direction.

We choose a characteristic length $L$ and a characteristic velocity
\begin{equation}
\label{bar:u}
   \bar{u} = \frac{\eta v_z}{ \Big{(}- \frac{\partial p}{\partial z}\Big{)} L^2 } .
\end{equation}
We can then make the Stokes equation in its dimensionless form
\begin{equation}
  \label{Poisson}
  \frac{\partial ^2 \bar{u}}{\partial \bar{x}^2} + \frac{\partial ^2 \bar{u}}{\partial \bar{y}^2} = -1 .
\end{equation}
The boundary conditions are $\bar{u} = 0$ at the liquid-solid interface and $ \textbf{n} \cdot \nabla \bar{u} = 0$ at the free surface, where $ \textbf{n} $ is the normal vector of the meniscus surface. 

The friction coefficient $\xi$ obeys Darcy's law
\begin{equation}
\label{Darcy_change}
   \frac{ \partial p}{\partial z} = - \xi Q ,
\end{equation}
where
\begin{equation}
  \label{eq:Q}
  Q = \int \ud x \ud y \, v_z
\end{equation}
is the volume flux.

Combining Eqs. (\ref{bar:u}), (\ref{Darcy_change}), and (\ref{eq:Q}), we can get $\xi$ as
\begin{equation}
\label{friction}
   \xi = \frac{\eta}{L^4 \int \mathrm{d} \bar{x} \mathrm{d} \bar{y} \bar{u}} .
\end{equation}
We performed numerical calculation using the finite element method in Matlab and obtained the results of $\int \mathrm{d} \bar{x} \mathrm{d} \bar{y} \bar{u}$.

For the finger flow in square tubes, we choose the radius of the inscribed circle $r$[Fig. \ref{fig:square_finger_f_s}] as the characteristic length $L$, then we get
\begin{equation}
\label{B1}
   \xi (s<s^*) = 106.5 \frac{\eta}{r^4 }.
\end{equation}
Substitute the relation $\zeta(s) = \xi(s) S^2 = \xi(s) (4a^2)^2$ into the above equation, we get
\begin{equation}
  \zeta(s) = 106.5 \times 16(1-\pi/4)^2  s^{-2} \eta = B_1 s^{-2} \eta, \quad B_1 = 78.50 .
\end{equation}

%%%%%%%%%%%%%%%%%%%%%%%%%%%%%%%%%%%%%%%%%%%%%%%%%%%%%%%%%%%%%%
%\bibliography{/home/zhou/bib/polymer}
%\bibliography{/Users/zhouj2/bib/polymer}